\documentclass{elsarticle}

\usepackage{lineno,hyperref}
\usepackage{subcaption}
\usepackage{amsmath,amssymb}
\usepackage{xcolor}
\usepackage[utf8]{inputenc}
\DeclareUnicodeCharacter{03B5}{\ensuremath{\varepsilon}}
\DeclareUnicodeCharacter{03F5}{\ensuremath{\varepsilon}}
\modulolinenumbers[5]

\newcommand{\blue}[1]{\textcolor{blue}{#1}}
\newcommand{\red}[1]{\textcolor{red}{#1}}
\newcommand{\cb}{{\mbox{\boldmath$c$}}}
\newcommand{\xb}{{\mbox{\boldmath$x$}}}

\journal{Computers \& Fluids}

\begin{document}

\begin{frontmatter}

\title{Data-driven RANS closures for three-dimensional flows around bluff bodies}

\author{Jasper P. Huijing}
\author{Richard P. Dwight\corref{mycorrespondingauthor}}
\ead{r.p.dwight@tudelft.nl}
\author{Martin Schmelzer\corref{duumy}}
\cortext[mycorrespondingauthor]{Corresponding author}

\address{Aerodynamics Group, Faculty of Aerospace Engineering, TU Delft. \\Kluyverweg 2, 2629HT Delft, The Netherlands}

\begin{abstract}
In this short note we apply the recently proposed data-driven RANS closure modelling framework of {\it Schmelzer et al.\ (2020)} to fully three-dimensional, high Reynolds number flows: namely wall-mounted cubes and cuboids at $\mathrm{Re}=40,000$, and a cylinder at $\mathrm{Re}=140,000$.  For each flow, a new RANS closure is generated using sparse symbolic regression based on LES or DES reference data.  This new model is implemented in a CFD solver, and subsequently applied to prediction of the other flows.  We see consistent improvements compared to the baseline $k-\omega$ SST model in predictions of mean-velocity in complete flow domain.
\end{abstract}


\end{frontmatter}


\section{Introduction}
Reynolds averaged Navier-Stokes (RANS) models are notoriously inaccurate in the presence of massive flow separation, for example in the wake of bluff bodies.  The steady RANS paradigm -- of modelling fluctuations at all scales -- is fundamentally ill-suited to representing the complex unsteady dynamics in the wake of e.g.\ a cylinder or wall-mounted cube.  Nonetheless, in a wide variety of industrial applications (notably the automotive industry) it would be extremely valuable to have access to RANS closures that give reasonably accurate predictions in wakes.

Data-driven turbulence modelling uses data from high-fidelity simulations (LES, DNS) or experiments to semi-automatically derive new closure models, see the surveys~\cite{duraisamyTurbulenceModelingAge2019,xiaoQuantificationModelUncertainty2019}.  The methods of supervised machine-learning are used to represent and fit models, e.g.\ neural networks~\cite{lingEvaluationMachineLearning2015b}, random-forests~\cite{wangPhysicsinformedMachineLearning2017a,kaandorpDatadrivenModellingReynolds2020,luanInfluenceTurbulenceAnisotropy2020}, gene-expression programming~\cite{zhaoRANSTurbulenceModel2020}, and sparse symbolic regression~\cite{schmelzerDiscoveryAlgebraicReynoldsStress2020,steinerDatadrivenTurbulenceModeling2020}.  The latter two methods have the advantage of generating concise expressions for closure models, which can be inspected, analysed and implemented easily.

In this work we apply the Sparse Regression of Turbulence Anisotropy (SpaRTA) framework first developed by Schmelzer et al.\ \cite{schmelzerDiscoveryAlgebraicReynoldsStress2020}.  The models produced are explicit algebraic Reynolds-stress models, based on $k-\epsilon$ or $k-\omega$ closures, but correcting both the turbulence anisotropy, and the t.k.e.\ production.  The required correction fields are solved for by injecting DES, LES or DNS data into the RANS equations; and a model is obtained by regressing these corrections against mean-flow quantities available to RANS.  This model can then be applied to predict a flow for which no reference data is available.

Though the objective of predicting bluff body flows with RANS might seem optimistic, SpaRTA has already been demonstrated with success for flows with significant separation~\cite{zhangCustomizedDatadrivenRANS2020}.  Also note: our intention here is not to derive new {\it general purpose} closure models.  At a minimum that would require a more diverse set of training flows.  Rather, we wish to demonstrate that our framework has the capability of constructing RANS closures that generalize acceptably for massively detached flows.  As such they may be useful as components of larger, general purpose models.

While using the framework of \cite{schmelzerDiscoveryAlgebraicReynoldsStress2020} and \cite{zhangCustomizedDatadrivenRANS2020}, this work extends those articles in several ways.  We consider flows at significantly higher Reynolds numbers ($\mathrm{Re}=140,000$ vs $\mathrm{Re}\simeq 10,000$ in \cite{schmelzerDiscoveryAlgebraicReynoldsStress2020}); in 3D vs only 2D previously; and we use DES data for the first time.  The 3D LES data source means that optimization of the symbolic regression for large data-sets is required, and we introduce a practical technique for library reduction.


\section{Methodology: SpaRTA}
The objective is to generate a RANS closure from reference data, that improves predictions for some class of flows.  For extended details of our methodology, see \cite{schmelzerDiscoveryAlgebraicReynoldsStress2020}.  In brief: the incompressible RANS $k-\omega$ SST equations are modified with correction terms $\blue{b^\Delta}$ and $\red{R}$ to:
\begin{align}
    U_j \partial_j U_i &= \partial_j \left[ -\frac{1}{\rho} P + \nu \partial_j U_i + \nu_T \partial_j U_i - 2k\blue{b^\Delta_{ij}}, \right], \notag\\
    U_j \partial_j k &=P_k + \red{R} -\beta^* \omega k + \partial_j \left[\left(\nu + \sigma_k \nu_t \right) \partial_j k \right], \notag\\
    U_j\partial_j \omega &= \frac{\gamma}{\nu_t} \big(P_k+\red{R} \big)-\beta \omega^2 +\partial_j \big[ \big(\nu+\sigma_{\omega}\nu_t \big) \partial_j\omega \big] \notag\\ &+ CD_{k\omega},
\label{eq:1}
\end{align}
with $\rho$ the fluid density, $U$, $P$ the mean velocity and generalized pressure, $\nu$, $\nu_T = \nu_T(k, \omega)$ the molecular- and turbulence viscosity, $k$, $P_k=-2k(b_{ij}+\blue{b_{ij}^{\Delta}})\partial_j U_i$ the turbulence kinetic energy and its production rate, and $\omega$ the specific dissipation rate.  See \cite{menterTwoequationEddyviscosityTurbulence1994} for details of remaining terms and coefficients.  The baseline SST model uses the Boussinesq assumption as a fundamental modelling premise, namely
\[
b_{ij} \simeq -\frac{\nu_t}{k}S_{ij},\quad S_{ij} := \frac{1}{2}(\partial_i U_j + \partial_j U_i).
\]
where $b_{ij}$ is the normalized turbulence anisotropy.  The purpose of $\blue{b^\Delta}$ is to relax this assumption by allowing for deviations from Boussinesq.  Similarly the SST $k$-equation is a model for the true t.k.e. equation; and the new term $\red{R}$ is placed to allow for modelling errors here.  In practice $\red{R}$ allows for control of turbulence intensity, and $\blue{b^\Delta}$ for control of turbulence anisotropy.

\paragraph{Solving for corrective fields}
Given full-field LES or DES data for flow A, our preliminary objective is to find corrective fields $\red{R(\xb)}$, $\blue{b^\Delta(\xb)}$ -- i.e.\ as functions of the spatial coordinates $\xb := (x,y,z)$ -- such that when the system \eqref{eq:1} is solved for the same flow A, the resulting $U$ and $k$ correspond well to LES/DES mean values.  This is achieved by injecting frozen LES/DES quantities into \eqref{eq:1}, and solving for the remaining unknowns $\red{R}$, $\blue{b^\Delta}$ and $\omega$. This procedure is named $k$-corrective-frozen-RANS~\cite{schmelzerDiscoveryAlgebraicReynoldsStress2020} and can be seen as a generalization of the ``frozen'' method for estimating t.k.e.\ dissipation rate from LES data~\cite{weatherittDevelopmentAlgebraicStress2017}.  We use DES/LES estimates of the Reynolds stress tensor and t.k.e. which include contributions from both modelled and resolved stresses.  Note that we deliberately conflate the $k$ of the $k-\omega$ model, and the true t.k.e.\ at this step.  The objective is to obtain a system of equations which is predictive of true t.k.e., at the cost of perhaps larger correction terms.

\paragraph{Model Regression}
To make predictions it is necessary to generalize these corrective fields by building a closure model.  We achieve this by now finding expressions for $\red{R(S,\Omega)}$, $\blue{b^\Delta(S,\Omega)}$ as functions of the mean strain-rate tensor $S$ and rotation rate tensor $\Omega_{ij} := \frac{1}{2}(\partial_j U_i - \partial_i U_j)$; quantities available to the RANS solver.  In particular we follow Pope's integrity basis formulation~\cite{popeTurbulentFlows2000b}, which specifies that the most general functional form (under modest assumptions) for $b^\Delta(S_{ij},\Omega_{ij})$ can be written:
\begin{equation}
    b^\Delta_{ij}(S_{ij},\Omega_{ij})=\sum_{k=1}^{10} T_{ij}^{(k)}\alpha_k(\lambda_1,...,\lambda_5),
\end{equation}
where $T^{(1)} := S$, $T^{(2)} := S\Omega -\Omega S$, etc.\ are basis tensors, $\lambda_l$ are the five invariants of $S$ and $\Omega$~\cite{johnsonHandbookFluidDynamics2016}, and $\alpha_k:\mathbb{R}^5\rightarrow \mathbb{R}$ are ten, arbitrary scalar-valued functions.  By this construction $b^\Delta$ is always symmetric and traceless, the map $b^\Delta(\cdot)$ is invariant under rotations, and Galilean invariant by virtue of the dependence only on $\partial_jU_i$.  In this work we use only $T^{(1)}, \dots, T^{(4)}$ and $\lambda_1$, $\lambda_2$, thereby restricting our search to {\it quadratic} nonlinear eddy-viscosity models, following e.g.~\cite{spezialeNonlinearKlKe1987,shihNewEddyViscosity1995}.  

Aiming for simple algebraic expressions, we represent each $\alpha_k(\lambda_1,\lambda_2)$ as a linear combination of a large {\it library} of basis functions $\mathcal{L} = (\phi^1, \dots,\phi^L)$:
\begin{equation}
\alpha_k(\lambda_1,\lambda_2) = \sum_{l=0}^L c_k^l \phi^l(\lambda_1,\lambda_2), \quad k\in\{1,\dots,4\},
\label{eq:3}
\end{equation}
where the library is generated from products and powers of the inputs $\mathcal{L} = (1,\lambda_1,\lambda_1^2,\dots,\lambda_2,\lambda_2^2,\dots,\lambda_1\lambda_2,\dots)$, see for example~\cite{kornsAbstractExpressionGrammar2011}.  To avoid models with large numbers of terms, as well as overfitting, we apply elastic net regression~\cite{zouRegularizationVariableSelection2005}, which encourages sparsity (most coefficients $c_k^l = 0$), to find the model form.  Let $\cb\in\mathbb{R}^{4\times L}$ represent the vector of all model coefficients, then we solve
\begin{equation}
\min_\cb \|b^\Delta(\xb) - b^\Delta(\cb;S(\xb),\Omega(\xb)) \|^2 + \theta\rho\|\cb\|_1 + \theta(1-\rho)\|\cb\|_2^2,
\label{eq:2}
\end{equation}
where in practice the first norm is estimated using the points of the mesh used to obtain the corrective field $b^\Delta(\xb)$.  The term $\|\cb\|_1:=\sum |c_i|$ encourages sparsity of $\cb$ and the term $\|\cb\|_2^2:=\sum c_i^2$ controls the magnitude of nonzero coefficients. Both are blended by $\rho\in[0,1]$ and $\theta\in\mathbb{R}^+$ ultimately controls the extent of regularization.  Using path elastic net~\cite{friedmanRegularizationPathsGeneralized2010}, a large number of candidate models for various $\rho$, $\theta$ are obtained, and a second ridge-regression step is used to select coefficients.  A similar procedure is applied to model $R(S,\Omega)$: we assume the form $R = 2k\partial_j U_i \hat R_{ij}(S,\Omega)$, and use the base-tensor series to model $\hat R_{ij}$.

With the extension from 2D cases in \cite{schmelzerDiscoveryAlgebraicReynoldsStress2020} to 3D cases here the size of the data-set in \eqref{eq:2} has increased significantly, and symbolic regression becomes a significant memory bottleneck due to storage of the library $\mathcal{L}$ evaluated on the full data-set.  We introduce a {\it cliqueing} procedure motivated by the high multi-collinearity observed in $\mathcal{L}$.  Specifically we compute the correlation coefficient between all pairs of library functions $\phi^i$, and sort them into {\it cliques} whose correlation within a clique always exceeds a cut-off (of $0.99$).  Efficiently finding cliques is an established problem in graph theory~\cite{albaGraphTheoreticDefinition1973}.  We then select the algebraically simplest member of the clique to represent the clique, and discard the remainder.   Although $\phi^i$ are nonlinear in the inputs, a linear measure of correlation is adequate, as they are combined linearly in \eqref{eq:3}.  This method is reminiscent of {\it elite basis regression}~\cite{Chen2017}, except that we are not concerned with correlation with the target, only with basis functions amongst themselves.

\section{Results}
We examine three flows: a wall-mounted cube in a channel (Flow A), a wall-mounted cuboid (length:width:height ratio 3:2:2) also in a channel (Flow B), and an infinite circular cylinder (Flow C).  Flows A, B are at $\mathrm{Re}=40,000$ based on bulk velocity and cube/cuboid height $h$, with the channel of height $2h$, and are well-used experimental~\cite{martinuzziFlowSurfaceMountedPrismatic1993} and numerical test-cases~\cite{alfonsiCoherentStructuresFlow2003}.  Flow C is at $\mathrm{Re}=140,000$ based on cylinder diameter~\cite{breuerChallengingTestCase2000}.  All three flows include massive separation, resulting in unsteady wakes with a wide range of time- and length-scales.

Ground-truth reference fields for Flows A, B are obtained with DES~\cite{spalartNewVersionDetachededdy2006} simulating a domain of size $14.5h \times 9h \times 2h$, the obstacle centred at $x=4h$, with periodic boundary-conditions in the cross-channel direction, and synthetic channel turbulence at the inflow plane~\cite{polettoNewDivergenceFree2013}, and an averaging time of $\simeq 12.4$ flow-throughs.  The RANS part of the DES is based on the Spalart-Allmaras one-equation model~\cite{spalartOneequationTurbulenceModel1992}, in contrast to our use of $k-\omega$ SST in our enhanced RANS.  

For Flow C ground-truth comes from a wall-resolved LES.  The cylinder has diameter $d$ and simulated length $\pi\cdot d$ with periodic boundaries.  The flow is fully turbulent, with synthetic turbulence at the inlet plane.  For all flows, velocity profiles were compared with existing published results, and found to be sufficiently accurate for the application.

\begin{figure}
    \centering
    \begin{subfigure}[b]{0.9\textwidth}
        \includegraphics[width=\textwidth]{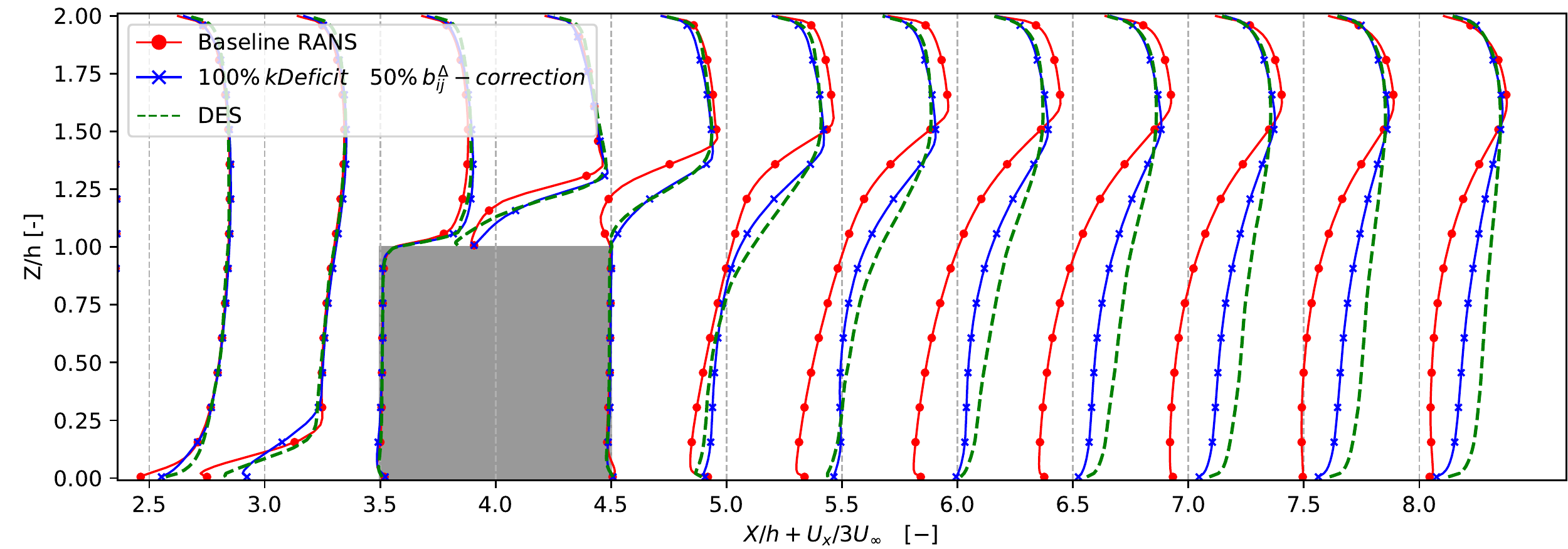}
        \caption{Centre plane $y=4.5h$}
        \label{fig:cube_y_Ux}
    \end{subfigure}
    \begin{subfigure}[b]{0.9\textwidth}
        \includegraphics[width=\textwidth]{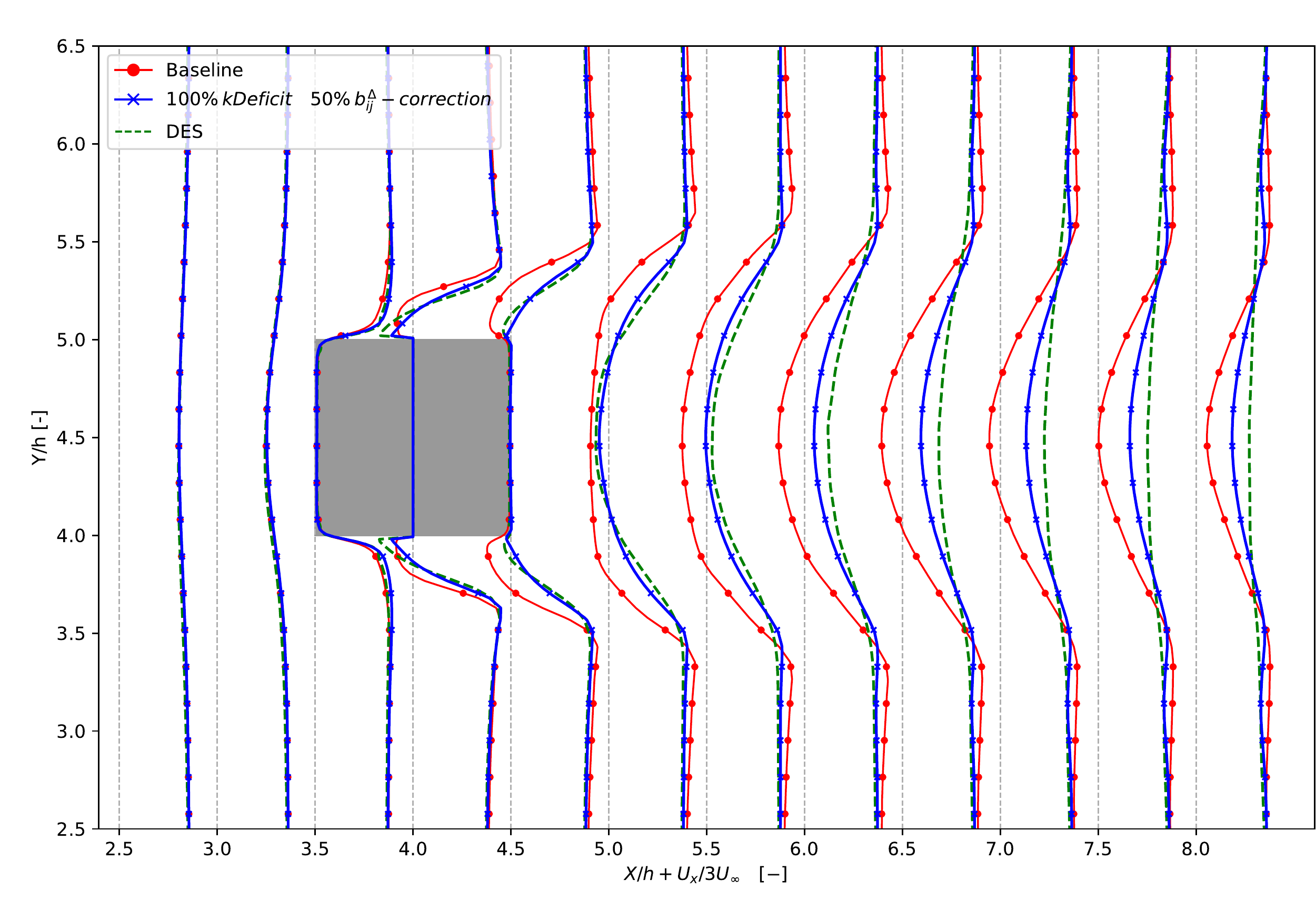}
        \caption{Mid-height of the cube $z=0.5h$.}
        \label{fig:e_cube_z_Ux}
    \end{subfigure}
    \caption{Flow A: x-component velocity comparison.}
    \label{fig:cube}
\end{figure}

\paragraph{The ``Cube'' model}
We first apply the SpaRTA methodology to Flow A.  Mean velocity profiles for the DES reference and baseline RANS are shown in Figure~\ref{fig:cube}.  The baseline significantly overestimates the size of the recirculation region on the top and sides of the cube, and the wake recovery is very significantly delayed.  Frozen corrective fields are obtained, and symbolic regression is used to reduce these to a closure model, giving the following correction to $k-\omega$ SST:
\begin{equation*}
    \begin{split}
        b_\mathrm{Cube}^{\Delta}&=T^{(1)}(28.68 \lambda_{1} + 4.717 \lambda_{2} - 0.3560)\\
        &+ T^{(2)}(-88.99 \lambda_{1} + 68.77 \lambda_{2} + 13.80)\\
        &+ T^{(3)}(20.59 \lambda_{1} - 0.8594) \\
        &+ T^{(4)}(11.46 \lambda_{1} - 2.770 \lambda_{2} - 0.94412),
    \end{split}
    \label{eq:correction_model_cube_bij}
\end{equation*}
\begin{align*}
      \hat R_\mathrm{Cube} &= T^{(1)}(-35.74 \lambda_{1} \lambda_{2} - 69.58 \lambda_{1} \\
      &\qquad\quad+ 39.74 \lambda_{2}^2 + 7.573 \lambda_{2} + 3.739)\\
    &+ T^{(3)}(5.867 \lambda_{1} + 0.3755\lambda_{2}^2 - 6.784 \lambda_{2} - 16.48)\\
    &+ T^{(4)}(-2.304 \lambda_{1} - 0.2182\lambda_{2}^2 + 3.136 \lambda_{2} - 4.822).
\label{eq:correction_model_cube_R}
\end{align*}
These corrections are implemented in our solver, a modified version of \texttt{simpleFoam} in OpenFOAM, as a new turbulence model -- this is straightforward and can be completely automated.  The RANS solver is run with the new model for Flow A as a verification check.  The resulting mean-flow is also shown in Figure~\ref{fig:cube}, where the flow in the entire domain is seen to better represent the reference.  In particular the flow near the stagnation point, the separation on the top and sides, and the wake recovery are all closer to the reference -- although wake recovery is still somewhat under-predicted.

Note that to obtain these results it was necessary to relax $b_\mathrm{Cube}^{\Delta}$ by $50\%$ to gain stability.  In general we observed that the $\red{R}$ correction term acted to increase the production of $k$, while the contribution of $\blue{b^\Delta}$ to the production term was generally negative in the wake.  In the frozen corrective fields these effects cancelled to some extent, but the same cancellation was less apparent in the discovered models.  As it is, the relaxation of $\blue{b^\Delta}$ by $0.5$ will tend to increase the eddy-viscosity beyond that recommended by the frozen corrective fields, and thereby stabilize the simulation.

\begin{figure}
    \centering
    \begin{subfigure}[b]{0.9\textwidth}
        \includegraphics[width=\textwidth]{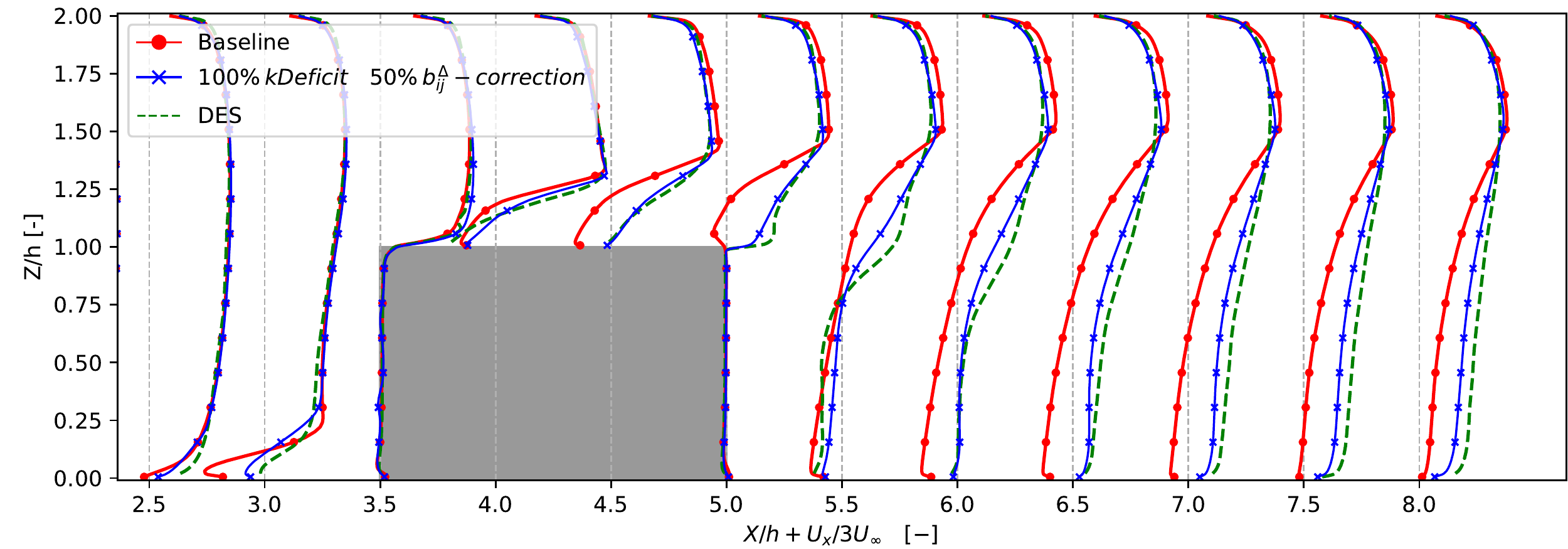}
        \caption{Centre plane $y=4.5h$}
        \label{fig:cube_extended_y_Ux}
    \end{subfigure}
    \begin{subfigure}[b]{0.9\textwidth}
        \includegraphics[width=\textwidth]{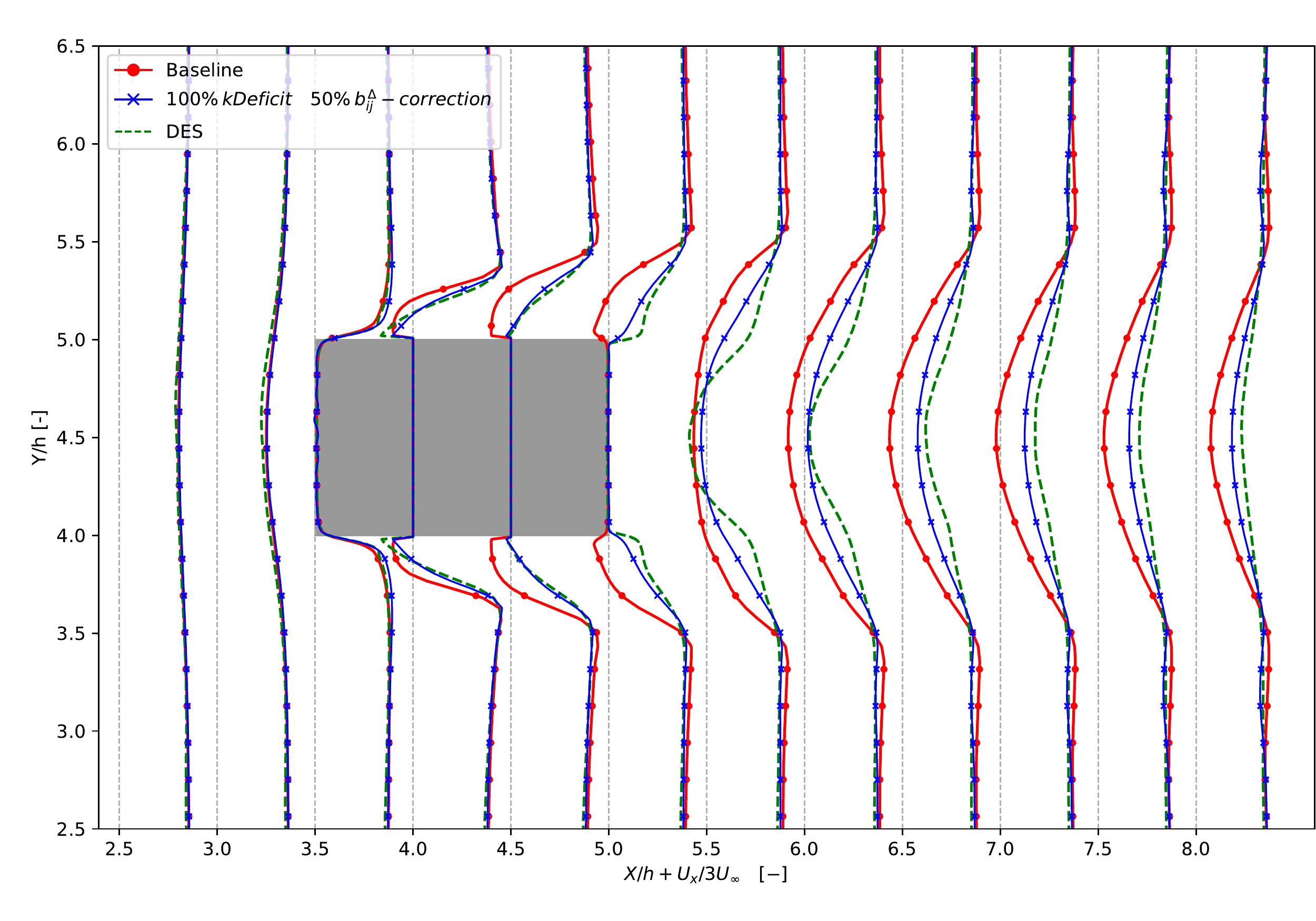}
        \caption{Mid-height $z=0.5h$}
        \label{fig:cube_extended_z_Ux}
    \end{subfigure}
    \caption{Flow B: Extended cube x-component velocity comparison.}
    \label{fig:cube_extended}
\end{figure}

To make a prediction, the ``Cube'' model is applied to Flow B.  The mean velocity profiles are shown in Figure \ref{fig:cube_extended}.  The flow is topologically similar to Flow A, as such it is reasonable to hope that the model generalizes well.  This is indeed the case, with significant improvements over the RANS baseline visible everywhere in the domain.  Notably, on the longer top and sides of the obstacle, the reproduction of the ground-truth is good -- while the flow separation is quite different.  A weakness can be seen in the near wake, where in Figure~\ref{fig:cube_extended_z_Ux} a much narrower shear-layer is visible in the reference compared to the corrected model, however the wake recovery is good.

As a second prediction, the Cube model is applied to Flow C, see Figure~\ref{eq:cylinder_bij}.  In this flow the baseline RANS predicts the flow upstream of the obstacle very well, and the correction model does not modify the flow here.  Once again however the baseline predicts much too slow wake recovery, and this is corrected almost completely by our model based only on data from Flow A.
\begin{figure}
    \centering
    \includegraphics[width=0.9\textwidth]{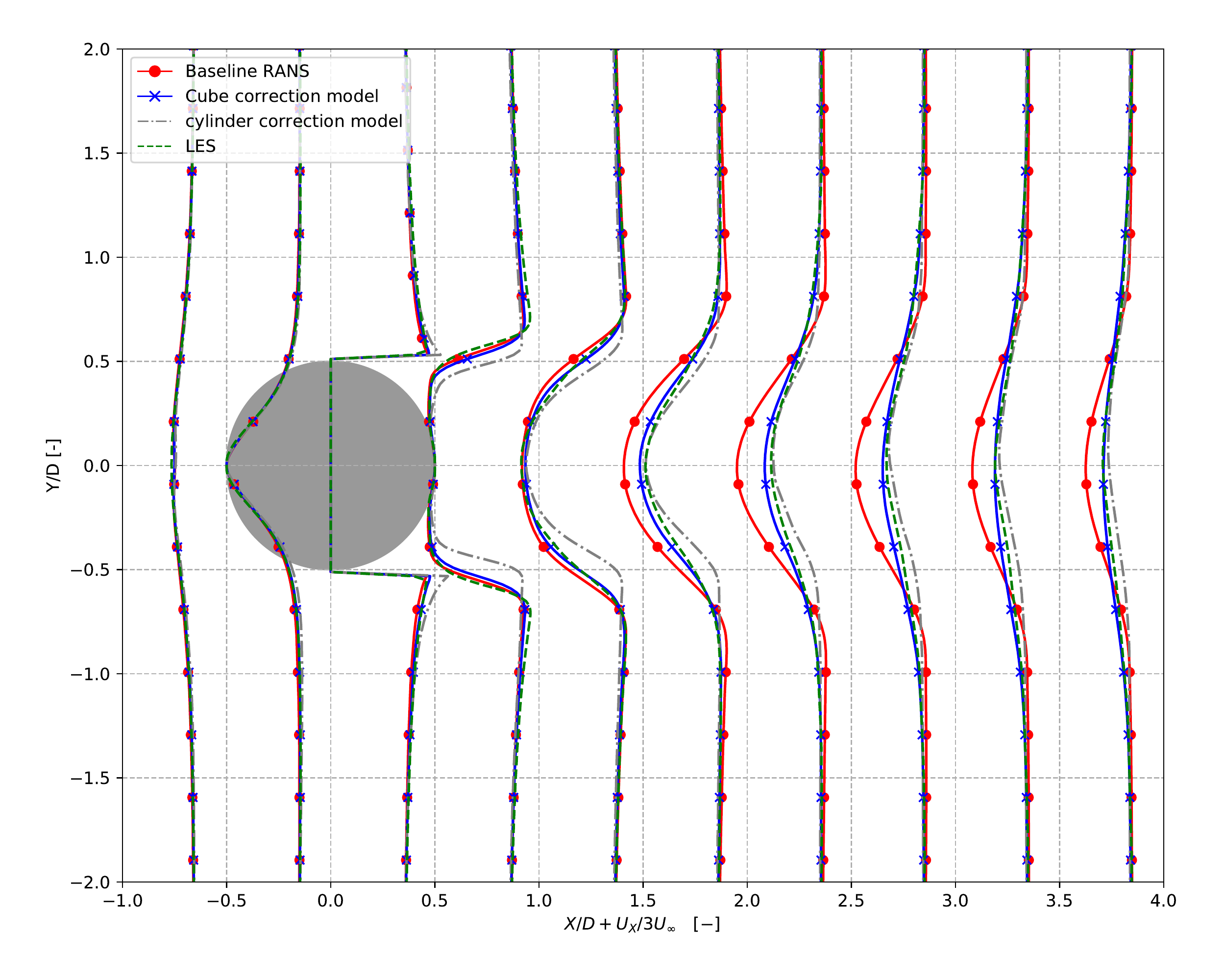}
    \caption{Flow C: x-component velocity comparison}
    \label{fig:cylinder_U_x}
\end{figure}

\paragraph{The ``Cylinder'' model}
To investigate the sensitivity of the data-source on the resulting model and its performance,
we apply the SpaRTA methodology using the wall-resolved LES data of Flow C.  Again the frozen correction fields are found readily, and the symbolic regression approach gives:
\begin{equation}
\begin{split}
    b_\mathrm{Cylinder}^{\Delta}&=
    T^{(1)}(19.24 \lambda_1 + 57.86 \lambda_2 + 2.939)\\
    &+ 9.695\cdot T^{(2)} + 7.805\cdot T^{(3)} + 1.171\cdot T^{(4)},
    \end{split}
    \label{eq:cylinder_bij}
\end{equation}
\begin{equation}
\begin{split}
    \hat R_\mathrm{Cylinder} &= T^{(1)} (-2.823 \lambda_{1} + 33.82 \lambda_{2} + 2.586)\\
    &+ 11.17\cdot T^{(3)} - 3.107\cdot T^{(4)}.
    \end{split}
    \label{eq:cylinder_R}
\end{equation}
In this case it was possible to find a significantly sparser model that nonetheless reduced the regression error to an acceptable level.  This is likely at least partially thanks to the two-dimensionality of the mean-flow.  Again the model was implemented in our OpenFOAM-based solver, and Flow C was predicted as a verification exercise -- see Figure~\ref{eq:cylinder_bij}.  Once more the flow is significantly more accurate than the baseline (and especially in terms of wake recovery) except in a small region in the near wake.  Notably the ``Cylinder'' model performs worse for this verification exercise, than the ``Cube'' model does in prediction.  We suspect this is partially a consequence of the relatively diverse flow content of the cube data-set -- the choice of training data will be a subject of future work.

\begin{figure}
    \centering
    \includegraphics[width=0.9\textwidth]{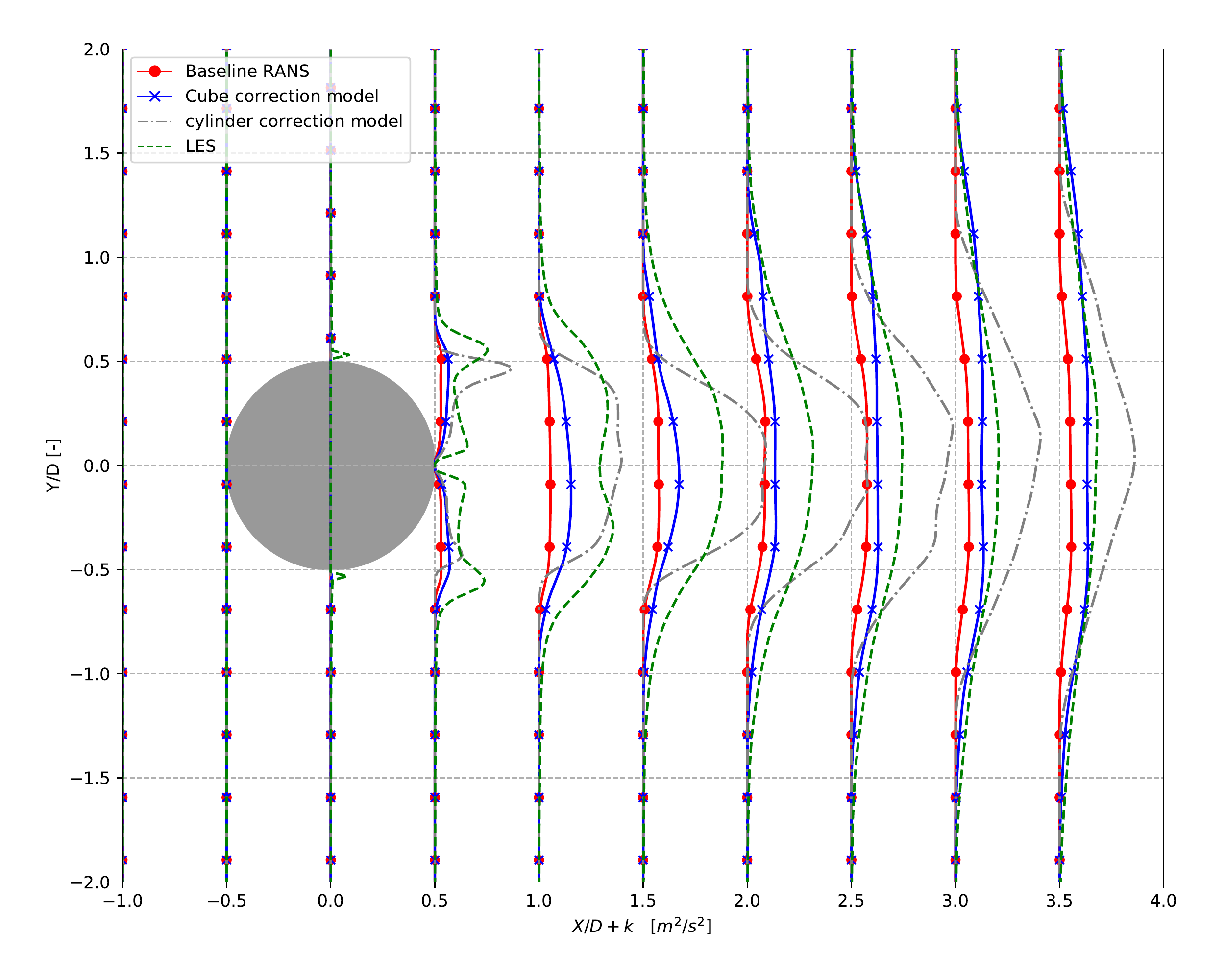}
    \caption{Flow C: Turbulent kinetic energy comparison}
    \label{fig:cylinder_k}
\end{figure}

Finally, in Figure \ref{fig:cylinder_k} $k$-profiles for Flow C are compared.  Again both correction models improve dramatically on the baseline -- which massively under-predicts $k$ in the wake -- however neither are particularly accurate.  A likely cause is the large-scale dynamics of the K\'arm\'an vortex street, which contribute to the LES estimate of t.k.e.\ but are missing in RANS.  Once more however the free-stream is unaltered.

\section{Conclusions}
We've demonstrated the ability to construct custom explicit algebraic Reynolds stress models for bluff-body flows from LES and DES data.  We've shown the models have a generalization capability, with respect to different geometries and different flows, while consistently outperforming baseline models.  This study shows the possibility of using RANS for massively separated flows using suitable machine-learning techniques, though many questions remain.  Specifically: Is the method robust to training-data choice?  Is symbolic regression general enough to capture the necessary corrections?  How do the $\blue{b^\Delta}$ and $\red{R}$ corrections interact in $k$-production?  Further work will focus on investigating the stability of automatically generated models, and scaling up training data from one flow, to a large database of flows.

\bibliography{turbml}

\end{document}